\documentclass[twocolumn,secnumarabic,amssymb, nobibnotes,nofootinbib, aps, prd]{revtex4-1}
\pdfoutput=1

\usepackage{graphicx}
\usepackage{enumitem}
\usepackage{latexsym}
\usepackage{amsfonts}
\usepackage{amssymb}
\usepackage{color}
\usepackage{amsmath}
\usepackage{slashed}
\usepackage{dcolumn}
\usepackage{verbatim}

\definecolor{darkblue}{rgb}{0,0,0.5}
\usepackage[
pdfauthor={Nirmal Raj}]{hyperref}

\DeclareGraphicsExtensions{.pdf,.png,.jpeg}

\definecolor{purple}{rgb}{1,0,1}

\newcommand{\beq}{\begin{equation}}
\newcommand{\eeq}{\end{equation}}
\newcommand{\bea}{\begin{eqnarray}}
\newcommand{\eea}{\end{eqnarray}}
\newcommand{\nn}{\nonumber}
\def\shat{\sqrt{s}}

\newcommand{\dm}{\chi}
\newcommand{\med}{\widetilde{T}}
\newcommand{\smed}{\widetilde{a}}
\newcommand{\decay}{\Gamma_{\rm MED}}
\newcommand{\mdm}{m_{\rm DM}}
\newcommand{\mmed}{m_{\rm MED}}
\newcommand{\msmed}{m_{\rm MED}}
\newcommand{\mtop}{m_t}
\newcommand{\deltild}{\widetilde{\delta}}
\newcommand{\ptild}{d^3\widetilde{p}}

\def\sigmaveeave{\langle \sigma v\rangle_{\rm ann}}
\def\sigmavee{\langle \sigma v\rangle}
\def\relabund{\Omega_\chi h^2}
\def\lag{\mathcal{L}}
\def\tchlabelf{{\tt tchFDM}}
\def\tchlabels{{\tt tchSDM}}
\newcommand{\met}{\slashed{E}_T}

\newcommand{\half}{\frac{1}{2}}

\newcommand{\lsim}{\lesssim}
\newcommand{\ra}{\rightarrow}

\newcommand{\ldm}{\lambda_\chi}
\newcommand{\lsm}{\lambda_t}

\allowdisplaybreaks

\begin{document}	

\title{
  Forbidden Dark Matter at the Weak Scale via the Top Portal \\}

\author{Antonio Delgado}
\author{Adam Martin}
\author{Nirmal Raj}
\affiliation{Department of Physics, University of Notre Dame, 225 Nieuwland Hall, Notre Dame, Indiana 46556, USA}

\begin{abstract}
 At the tail of its velocity distribution, cold dark matter (DM) can annihilate at finite temperature to states heavier than itself. 
We explore the possibility that DM freezeout is dictated by these
``forbidden annihilations" at the electroweak scale. 
Demanding that annihilation products be Standard Model particles, we find that for the forbidden mechanism to primarily set the
DM relic abundance, DM must couple predominantly, if not solely, to top quarks.
This can be arranged by invoking a non-trivial flavor structure such as Minimal Flavor Violation.
We avail two avenues to achieve the correct thermal cross-section, requiring a mediator exchanged in the $s$- or $t$-channel.
These simplified models submit easily to direct detection and collider searches, and necessarily hide from indirect detection signals.
Viable supersymmetric spectra involving the forbidden mechanism may be found if combined with co-annihilation. 
\end{abstract}

\maketitle

\section{Introduction}
\label{sec:intro}

Cosmological experiments indicate that about a quarter of the energy budget of the universe is sourced by dark matter (DM).
How did DM acquire this abundance?
The intimacy between thermodynamics and the history of the early universe suggests that it may have been set 
by some thermal mechanism. 
Usually, DM is assumed to have frozen out of equilibrium from a thermal bath of various particle species.
In the simplest models, the DM relic abundance $\relabund$ is set by a Lee-Weinberg mechanism \cite{Lee:1977ua} involving the process $\dm \dm \ra {\rm SM~SM'}$.
The key ingredient here is the thermally averaged  annihilation cross-section, $\sigmaveeave$.
To satisfy $\relabund$ measured by Planck \cite{Ade:2015xua}, one requires, at the stage of freeze-out,
\beq
\sigmaveeave = 3 \times 10^{-26}~{\rm cm}^3~{\rm s}^{-1} .
\label{eq:thermalxs}
\eeq
This value is obtainable in a minimal manner by parametrically taking $\sigmaveeave \sim \alpha^2/M^2$, 
where $\alpha$ and $M$ are the DM-Standard Model (SM) coupling and DM mass respectively.
Strikingly, electroweak-size couplings and weak scale masses -- a combination generically required to understand the stability of the Fermi scale -- can lead to Eq.~(\ref{eq:thermalxs}), a coincidence denoted in the literature as the ``WIMP miracle".
However, there are well-known exceptions to this possibility.
These alternative mechanisms for obtaining the correct $\sigmaveeave$ have gained traction in recent decades in light of stringent limits placed on the minimal WIMP scenario.
Let us briefly review the mechanisms prominent in the literature:

\begin{itemize}
\item The combination $\alpha^2/M^2$ may be fixed by a choice of coupling strength and mass that need not be related to the weak scale, the so-called {\bf WIMPless miracle} \cite{Feng:2008ya}.

\item The usual Lee-Weinberg calculation does not apply to the ``{\bf semi-annihilation}" process  $\dm_i \dm_j \ra \dm_k {\rm S}$, where $S$ is a singlet under DM symmetries. 
This process must be included in the computation of freezeout if allowed by the model \cite{D'Eramo:2010ep}.

\item $3 \to 2$ processes within the dark sector, as opposed to $2 \to 2$ DM annihilation to SM particles, may be the dominant DM number-changing process.
This occurs for {\bf SIMP DM}~\cite{Hochberg:2014dra, Hochberg:2014kqa}.
These scenarios require a DM-SM connection in order to be viable and, unlike the other exceptions listed here, DM in this case is automatically self-interacting.
 
\item An effective $\sigmaveeave$ that enters the computation of DM relic density may be set by multiple processes.
In the mechanism of {\bf co-annihilation}, all possible annihilation combinations between DM and nearly mass-degenerate states must be included \cite{Griest:1990kh,Gondolo:1990dk}.
 
\item In $s-$channel-mediated annihilations, the amplitude increases near a pole, enhancing $\sigmaveeave \sim |\mathcal{M}|^2$.
This is the so-called {\bf resonant annihilation} \cite{Griest:1990kh,Gondolo:1990dk}.

\item The thermal averaging of the annihilation cross-section may play a non-trivial role \cite{Griest:1990kh,Gondolo:1990dk}.
In particular, if the  products of DM annihilation are heavier than DM, their limited phase space truncates the range of DM velocites over which the averaging is performed. 
In such cases, the required value of $\sigmaveeave$ can be achieved if the non-thermal cross-section is large to begin with.
Since this type of annihilation cannot proceed at zero temperature, this class of models is referred to as ``{\bf forbidden dark matter}".
\end{itemize}

Currently, extensive DM searches through direct and indirect detection, and collider production, target the electroweak scale. 
These searches are usually complementary and capable of probing the mechanism underlying DM freezeout.
While considerable attention has been paid to phenomenological implications of semi-annihilation, co-annihilation and resonant annihilation at the weak scale, the same cannot be said for forbidden annihilation.
This is surprising, given that DM annihilation to heavier states is a minimal deviation from standard WIMP freezeout.
In the quest of demystifying the nature of DM,
it is imperative to explore the full landscape of abundance-setting mechanisms. 
The viability of the forbidden mechanism at the electroweak scale, a scenario that may be loosely described as forbidden WIMPs, is the primary concern of this paper.

The rest of the paper is set up as follows.
Sec.~\ref{sec:model-building} first reviews the literature of forbidden DM models in order to distinguish our work.
It then deals with aspects of building forbidden WIMP simplified models and shows that the top quark portal is the only viable model if we require $\relabund$ to be set primarily by a forbidden mechanism.
Constraints on top portal models and future search sensitivities are studied in Sec.~\ref{sec:topportal}. 
Sec.~\ref{sec:discs} provides some discussions and concludes the paper.

\section{Model-Building Aspects}
\label{sec:model-building}

\subsection{Literature Review}
\label{subsec:litreview}

{We begin by pointing out the principal difference between previous work on forbidden DM and our paper. 
Whereas previous authors targeted indirect detection as the main means of probing their models,
our models cater exclusively to colliders and direct detection experiments.
In the following we will expand on this statement.

Forbidden DM was studied in \cite{Jackson:2009kg,Jackson:2013pjq,Jackson:2013rqp}
in the context of line signals in the sky.
In this series of papers, the SM gauge group was extended to include a $U(1)'$ sector, with its gauge boson $Z'$ mediating DM-SM interactions through interactions with the top.
For DM lighter than the $Z'$ and top quark, annihilation into the corresponding channels is suppressed, in turn quelling continuum photon emission.
This gives way to the domination of line signals produced by, e.g., $\gamma \gamma, \gamma h, \gamma Z$ final states from diagrams involving loops of the top quark and/or new fermion partners.  
When the DM relic density was computed, it was by taking these final states into account; the contribution from forbidden final states was not included.

Forbidden DM was invoked again by the authors of \cite{Tulin:2012uq} to explain the (former) anomaly of the 130 GeV Fermi line.
A new vector-like fermion with mass $\geq 130$~GeV was introduced as the state to which DM annihilated, and a singlet pseudoscalar played the role of mediator.
The relic density was not considered.
 
More recently, Ref.~\cite{D'Agnolo:2015koa} re-introduced forbidden DM into the literature; here, the final state of annihilation is a massive dark photon, the gauge mediator of a hidden $U(1)'$ sector.
Emphasis was laid on a new calculation of the relic density.
Constraints on this model arise from possible gauge kinetic mixing, mainly from beam dump experiments and observations of the CMB and supernovae cooling.

Our approach to forbidden DM will differ from these models in several important respects:
\begin{itemize}[label={}]
\item(i) In keeping with minimality, we do not extend the SM gauge group.
Instead, we consider ``simplified models", effective low-energy theories that capture DM-SM interactions via dimension-4 Lagrangian terms 
\cite{Chang:2013oia,
An:2013xka,
Bai:2013iqa,
DiFranzo:2013vra,
Papucci:2014iwa,
Garny:2014waa,
Abdallah:2014hon,
Buckley:2014fba,
Harris:2014hga,
Abdallah:2015ter,
Altmannshofer:2014cla,
Baker:2015qna}.
In these models the new physics sector usually comprises of no more than the DM state and a mediator. 
The DM is charged odd under a $Z_2$ parity to ensure its stability.
Crucially, these models are highly sensitive to collider and direct detection experiments through the couplings of the mediator.
For a recent review of simplified models vis-a-vis LHC searches, see \cite{Abdallah:2015ter}.

We will study two different classes of simplified models, involving mediators either in the $s$- or $t$-channel.
These models are described in full detail in Sec.~\ref{sec:topportal}.

\item (ii) The DM abundance in our framework is set by forbidden annihilation to Standard Model states. 
In contrast, the abundance in \cite{Jackson:2009kg,Jackson:2013pjq,Jackson:2013rqp} was set by allowed annihilations to (lighter-than-DM) SM states.
In the instances where forbidden annihilation was considered, the abundance was not set to the observed value.
In ~\cite{Tulin:2012uq} and \cite{D'Agnolo:2015koa}, the annihilation products were exotic particles.

Enforcing the condition that DM annihilate to SM has two virtues. 
Since all the SM masses are known, we are able to obtain a fair indication of where the DM mass might lie.
Secondly, by not allowing DM to annihilate to mediator final states, we insure communication between the dark and luminous sectors, which is crucial for probing the forbidden mechanism under the lampposts of LHC and direct detection.

\item (iii) 
Thermally averaged forbidden cross-sections are evaluated by integrating over the tail of the DM velocity distribution, usually assumed Maxwell-Boltzmann.
Parametrically, $\sigmaveeave \sim \sigma_{\rm ann} \exp(- k \Delta M/T)$, where $\sigma_{\rm ann}$ is the unaveraged cross-section, $\Delta M$ the mass deficit of DM w.r.t. the final state and $k$ a constant.
In order to obtain Eq.~(\ref{eq:thermalxs}), a large $\sigma_{\rm ann}$ is required to overcome the exponential Boltzmann suppression.
This was achieved in \cite{Jackson:2009kg,Jackson:2013pjq,Jackson:2013rqp,Tulin:2012uq} with an $s$-channel mediator so that the annihilation could transpire near a pole, bolstering $\sigma_{\rm ann}$. 
In \cite{D'Agnolo:2015koa}, $\sigma_{\rm ann} \sim \alpha^2/M^2$ was enhanced by diminishing the mass scale of the dark system  $M$. 
A mediator lighter than 10 GeV was considered, with focus on the MeV scale.

In our approach, we will tackle this issue by two means. 
First, in simplified models with an $s$-channel mediator, annihilation near a pole can overcome Boltzmann suppression {\em \`a la} \cite{Tulin:2012uq}.
Second, in simplified models with a $t$-channel mediator, where the interactions take the form $\lag \supset \lambda \rm{(SM)(DM)(mediator)}$, the coupling $\lambda$ can be large ($>$ 1) as long as it is perturbative.
This is legitimate since simplified models are explicitly constructed as effective low-energy theories; in fact, large couplings were demonstrated to be a \textit{requisite} for obtaining the correct relic density \cite{Chang:2013oia,An:2013xka,Altmannshofer:2014cla} and for saturating collider bounds \cite{DiFranzo:2013vra}.
Sizeable couplings can then make the annihilation overcome Boltzmann suppression.
This amounts to saying that while \cite{D'Agnolo:2015koa} increased the ratio $\alpha^2/M^2$ by lowering $M$, it is here accomplished by raising $\alpha$.
\end{itemize}

Traditionally, $\sigmaveeave$ in the forbidden set-up is computed by weighting $\sigma_{\rm ann}$
with the Maxwell-Boltzmann distribution and carrying out a numerical integration over $v$ from the critical velocity to $\infty$. 
As earlier mentioned, Ref.~\cite{D'Agnolo:2015koa} proposed a new algorithm, taking advantage of the principles of unitarity and detailed balance.
By this method, the forbidden cross-section may be obtained by computing the cross-section of the reverse process, SM~SM~$\ra$~DM~DM.
We will adopt it for obtaining $\relabund$ throughout our paper. 
The details of the computation are enlarged in Appendix \ref{app:relabundcalc}.

\subsection{Simplified Forbidden WIMPs}
\label{sec:simpmods}

We now turn to building simplified models for forbidden WIMPs.
In principle, DM annihilation could proceed via some combination of forbidden and allowed processes.
That is, the sum of the final state masses could be greater or smaller than the sum of the DM masses.
In this paper, we will restrict ourselves to a \textit{principally} forbidden WIMP, a WIMP whose allowed annihilation channels are negligible.
We say ``principally", because the latter channels will always exist -- annihilation to light states through loops or three-body annihilations are usually unavoidable, though suppressed.
Our object here is construct a WIMP that is as forbidden as possible.
We do this to demonstrate that experiments are yet to rule out the forbidden mechanism as the chief means of providing the DM abundance\footnote{Forbidden WIMPs -- aren't.}.
The constraints one can impose on such a scenario then give a qualitative picture of a case where forbidden channels dominate the annihilation of DM.
Since we demand SM annihilation channels, we will now work through every massive SM particle and check whether it qualifies as the portal to a principally forbidden WIMP. 

\begin{itemize}

\item
{\em Electroweak bosons}

It has been long known that a WIMP charged under the electroweak gauge group and interacting solely with the gauge bosons is ruled out by experiment unless it mixes with an SM singlet \cite{Cohen:2011ec}.
Even so, weak gauge boson final states are unsuitable for our purpose, on the following grounds.
Assuming renormalizable interactions, it is easily checked that any DM annihilation with $Z Z$ (and possible $W W$) final states must have annihilations to light fermions through an $s$-channel $Z$.\footnote{An even stronger argument can be given for non-renormalizable interactions.
For concreteness, assume a scalar DM field $\dm$ that couples to the $Z$ boson through the operator $c_{\dm Z} |\dm|^2 Z_{\mu \nu} Z^{\mu \nu}/\Lambda_{\rm new}^2$
and that, through some tuning, the analogous operator involving photons vanishes at the relevant scale. 
Assuming the coefficient $c_{\dm Z}$ is $\mathcal{O}(1)$,
the $\dm$-$\dm$-$Z$-$Z$ coupling ($\equiv \lambda_{\dm Z}$) goes as $p_Z^2/\Lambda_{\rm new}^2$, where $p_Z$  is the $Z$ momentum.
Since the DM mass must be $\mdm \lsim m_Z$ for viable forbidden annihilations, the $Z$ velocity at freezeout should approximate the DM velocity, $v \sim 0.3$.
Thus, $p_Z \simeq m_Z v \simeq 30$ GeV.
For the validity of the contact interaction, $\Lambda_{\rm new}$ must be separated from $m_Z$ by at least an order of magnitude, thus we have 
$\lambda_{\dm Z} < (30~{\rm GeV}/900~{\rm GeV})^2 \simeq 10^{-3}$.
From our usual intuition about WIMPs, that the correct abundance for weak scale particles is obtained with electroweak-size couplings, we may be certain that such a feeble coupling as $\lambda_{\dm Z} < 10^{-3}$ would, due to suppressed annihilation, overclose the universe.
If we demand forbidden annihilation, $\sigmaveeave$ is
only further suppressed. 
These arguments hold also for fermion DM.
The only remedy here is to boost $\sigmaveeave$ by allowing for annihilations to lighter final states such as the photon.}

 This makes electroweak bosons ill-suited to our ends.

\item{\em Higgs boson}

DM interacting with the SM via a Higgs portal are an attractive study for their phenomenological richness. 
For DM DM $\ra h h$ to proceed via a forbidden mechanism, we need the DM to be lighter than $m_h =$ 125 GeV by a few percent. 
In this range, however, Higgs portal models that do not violate parity are in tension with limits placed by LUX and PandaX-II on DM-nucleon scattering \cite{LopezHonorez:2012kv,Craig:2014lda}.
This is the main reason why we will not pursue forbidden WIMP model-building via the higgs portal.
A second reason may be given: as in the case of annihilation to weak bosons, one finds that an annihilation to Higgs bosons is always accompanied by $s$-channel Higgs-mediated annihilations  
to other SM states, in this case light fermions and (one or more off-shell) gauge bosons.
We remark however that these channels are still suppressed due to the smallness of the Higgs Yukawa couplings and phase space respectively. 

\item{\em Light fermions}

If we expect forbidden DM to annihilate to the bottom quark and/or lighter quarks and leptons, we expect
the mediator(s) to be in a mass range of at most $\mathcal{O}(10~\rm GeV)$ --
heavier mediators would unduly suppress the annihilation.
Moreover, to enhance the forbidden cross-section the DM-SM couplings needed are usually large.
For these mediator masses and couplings, $s$-channel mediators would have been observed as resonances in di-fermion final states at colliders.
Similarly, $t$-channel mediators of mass $\leq$ 104 GeV would have been seen at LEP searches \cite{LEPcino,LEPslep} in missing energy signatures.
The lack of both these signals disfavors the possibility of light fermions being our annihilation products.

One potential exception to this argument is a DM particle coupling to the right-handed bottom quark via a colored mediator heavier than $104$~GeV, evading LEP searches. 
We discuss this in more detail at the end of Sec.~\ref{subsec:tch}.
The parameter space where this scenario is viable is extremely small, and for that reason we do not study it in this work.

\item{\em Top quark}

At the moment, no searches have substantially excluded a top portal DM and its mediators.
Hence we can proceed to construct simplified models involving it, with the proviso that flavor violations are avoided.
Note that even now, as stated in the beginning of this section, we cannot expect the $t$-$\bar{t}$ final state to contribute $100 \%$ to the WIMP annihilation. 
Processes involving loops that give the final states $b$-$\bar{b}, Z h, \gamma \gamma$, gluon-gluon, etc. are always present.
However, we find these to be highly sub-dominant to the the $t$-$\bar{t}$ channel due to loop factors and/or momentum-dependent vertices.
 Three-body final states such as $tWb$ may also be at play \cite{Chen:1998dp,Hosotani:2009jk,Yaguna:2010hn,Jackson:2013rqp}, but are strongly phase-space suppressed for the DM-top mass splittings we consider.
We will illustrate this further in Sec.~\ref{subsec:3body}.  
 As we will discuss shortly, in models with $s$-channel mediation, the imposition of Minimal Flavor Violation (MFV) leads to light quark channels. 
These are nevertheless negligible in comparison to the top quark channel due to their small Yukawa couplings. 

\end{itemize}

\begin{figure*}
\begin{center}
\includegraphics[width=.7\textwidth]{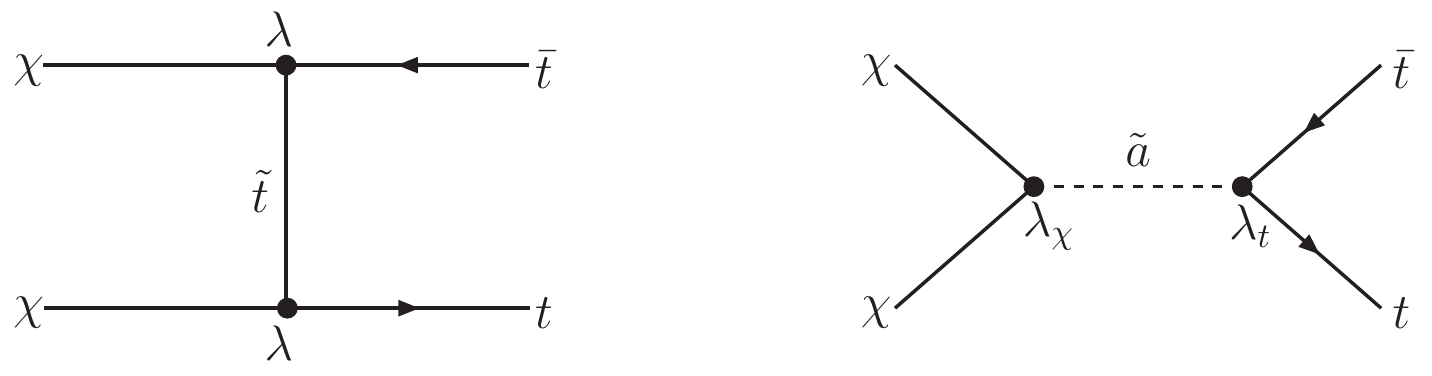}
\caption{
DM annihilation to top quarks can proceed via $s$-channel and $t$-channel processes.
For the $s$-channel the $\lsm$ vertex restricts $\smed$ to be spin-0, whereas $\dm$ can be spin-0 or spin-1/2, 
to avoid new sources of CP violation, we work with only spin-1/2 DM.
For the $t$-channel the spin of $\med$ (0 or 1/2) is fixed by that of $\dm$.
}
\label{fig:Feyn}
\end{center}
\end{figure*}

\section{Top Portal Forbidden WIMP}
\label{sec:topportal}

WIMPs can annihilate to top quark pairs via either an $s$-channel or $t$-channel mediator,
shown schematically in Fig.~\ref{fig:Feyn}.
The vertices in these diagrams capture all the interactions relevant to our phenomenology; 
the exact Lagrangians are provided in Appendix~\ref{app:formulae}.
In the t-channel case the DM particle $\dm$ could either be spin-0 or spin-1/2, and the mediator $\med$ is correspondingly spin-1/2 or spin-0.
On the other hand, the $s$-channel mediator
$\smed$ is always assumed spin-0 because it couples to SM fermion pairs.
We do not consider vector DM and mediators in this work, in keeping with our stance of not extending the SM gauge group.
A summary of the field content of our models is provided in Table~\ref{tab:fieldcontent}. 

\begin{table*}[hbt]
\begin{center}
\begin{tabular}{c c  c  c  c | c }
& & & \multicolumn{3}{c}{Spins} \\
\hline
Field  | &  $SU(3)_c \otimes SU(2)_W \otimes U(1)_Y$  & | $Z_2$ | & \multicolumn{2}{c}{$t$-channel}| &|  $s$-channel
\\
\hline \hline
$\dm$ & ({\bf 1,1,0}) & {\bf-1} & {\bf1/2} & {\bf0} & {\bf1/2}  \\ 
$\med$ &  ({\bf 3,1,2/3}) & {\bf-1} & {\bf0} &{\bf1/2} &\\ 
$\smed$ &  ({\bf 1,1,0}) & {\bf+1} &  &  & {\bf0} \\   \hline
&  & Free parameters |  & \multicolumn{2}{c}{$\lambda, \mdm, \mmed$}  | & | $\lsm, \ldm, \mdm, \mmed$ \\\end{tabular}
\end{center}
\caption{Summary of the field content of our models.}
\label{tab:fieldcontent}
\end{table*}

Of immediate concern at this point are constraints from flavor physics.
The fields introduced in these models can potentially induce dangerous flavor violating processes through loops.
To mitigate such effects, we invoke the principle of Minimal Flavor Violation (MFV) \cite{Hall:1990ac,Chivukula:1987py,Buras:2000dm,D'Ambrosio:2002ex} by which the flavor group is only broken by Yukawa spurions.
In the $s$-channel model, this means the singlet mediator couples to SM fermions proportional to their Higgs Yukawa strengths. 
The
couplings to the down and lepton sectors
 can be set to zero without offending MFV.
It was shown in Ref.~\cite{Dolan:2014ska} that even with MFV imposed, significant limits from flavor physics are incurred for DM lighter than 10 GeV.
This is not of concern for us, since our DM mass is a few GeV below the top quark.
In the $t$-channel model, one can imagine vertices analogous to those in Fig.~\ref{fig:Feyn} involving  two other generations of mediators.
Then MFV implies these couplings are flavor-diagonal such that every quark flavor communicates to $\dm$ via a unique mediator.
The effects of the ``up'' and ``charm" mediator can then be decoupled by setting them heavy.
Alternatively, flavor violation may be avoided in both models by assuming that the new physics couplings are aligned with the Yukawa matrices such that, in the mass basis, we are only left with couplings to the top quark.
The interesting possibility of DM, rather than the mediator, charged under the top flavor has been considered in Ref.~\cite{Kilic:2015vka}.
More general simplified models with DM carrying flavor indices are explored in Refs.~\cite{Kile:2011mn,Agrawal:2011ze}.

The key difference between the mediators in our models is their $Z_2$ charge.
From their interaction vertices in Fig.~\ref{fig:Feyn}, it can be seen that the $s$-channel ($t$-channel) mediator must be charged $Z_2 = +1 (-1)$. 
Hence $\smed$ can potentially mix with the SM Higgs boson while $\med$ has no mixing with the SM.
Notice also the difference in the number of free model parameters. 
In the first case, four are needed: the DM and mediator masses $\mdm$ and $\mmed$, and the DM-DM-mediator and top-antitop-mediator couplings $\ldm$ and $\lsm$.
The combination $\lambda \equiv \sqrt{\lsm \ldm}$ appears in all relevant cross-sections.
Moreover, the decay width of the mediator, $\decay$, usually plays a non-trivial role in $s$-channel annihilation and collider phenomenology.
Consequently the four free parameters are traded for $\mdm$, $\mmed$, $\lambda \equiv \sqrt{\ldm\lsm}$ and $\decay$. 
In $t$-channel mediation, three parameters describe the model: the DM and mediator masses $\mdm$ and $\mmed$, and the DM-top-mediator coupling strength $\lambda$.
The final difference between the scenarios is how they achieve the right relic abundance: 
as mentioned in Sec.~\ref{subsec:litreview}, forbidden annihilation is possible through the mediator $\smed$ due to cross-section enhancement near a pole, and through the mediator $\med$ due to putatively large $\lambda$.

On account of the differences detailed above, these two scenarios will result in very different phenomenologies. 
Hence we will consider each possibility separately in the following subsections.
Since our approach takes forbidden annihilation as the sole setter of DM abundance, we will fix $\lambda$ using the observed $\relabund = 0.1197$ when presenting our constraints.
This is done using the formulae presented in Appendices~\ref{app:relabundcalc} and \ref{app:formulae}, and comparing with {\tt MicrOmegas} 4.3 \cite{MuOmega} to double-check the result.
This method also allows all constraints to be shown in the space of mediator mass versus DM mass.
This is the treatment of simplified DM models adopted by \cite{Chang:2013oia}. 
In the following two sub-sections we will consider in more detail the $s$- and $t$-channel models, and the relevant phenomenology.

\begin{figure}
\begin{center}
\includegraphics[width=.45\textwidth]{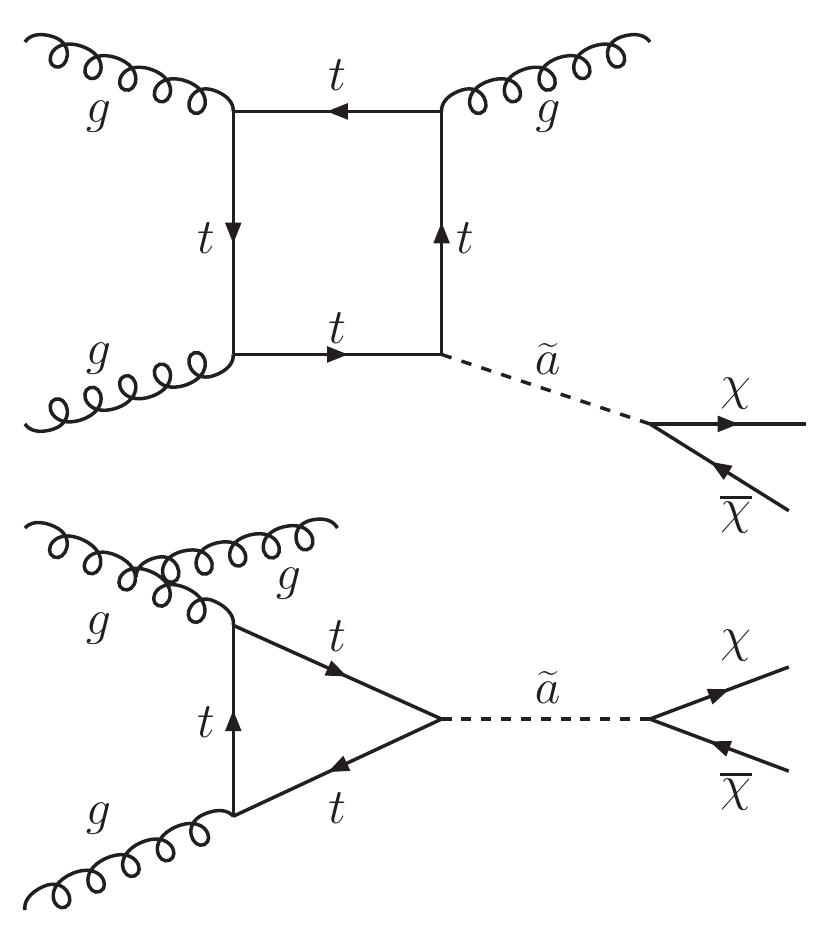}
\caption{
Example diagrams contributing to the monojet + MET signal in the $s$-channel model. 
For a CP-odd $\smed$, this cross-section is always higher than other channels like $t$-$\bar{t}$ and $b$-$\bar{b}$, yielding stricter bounds.} 
\label{fig:schFeyns}
\end{center}
\end{figure}

\begin{figure}
\begin{center}
\includegraphics[width=.45\textwidth]{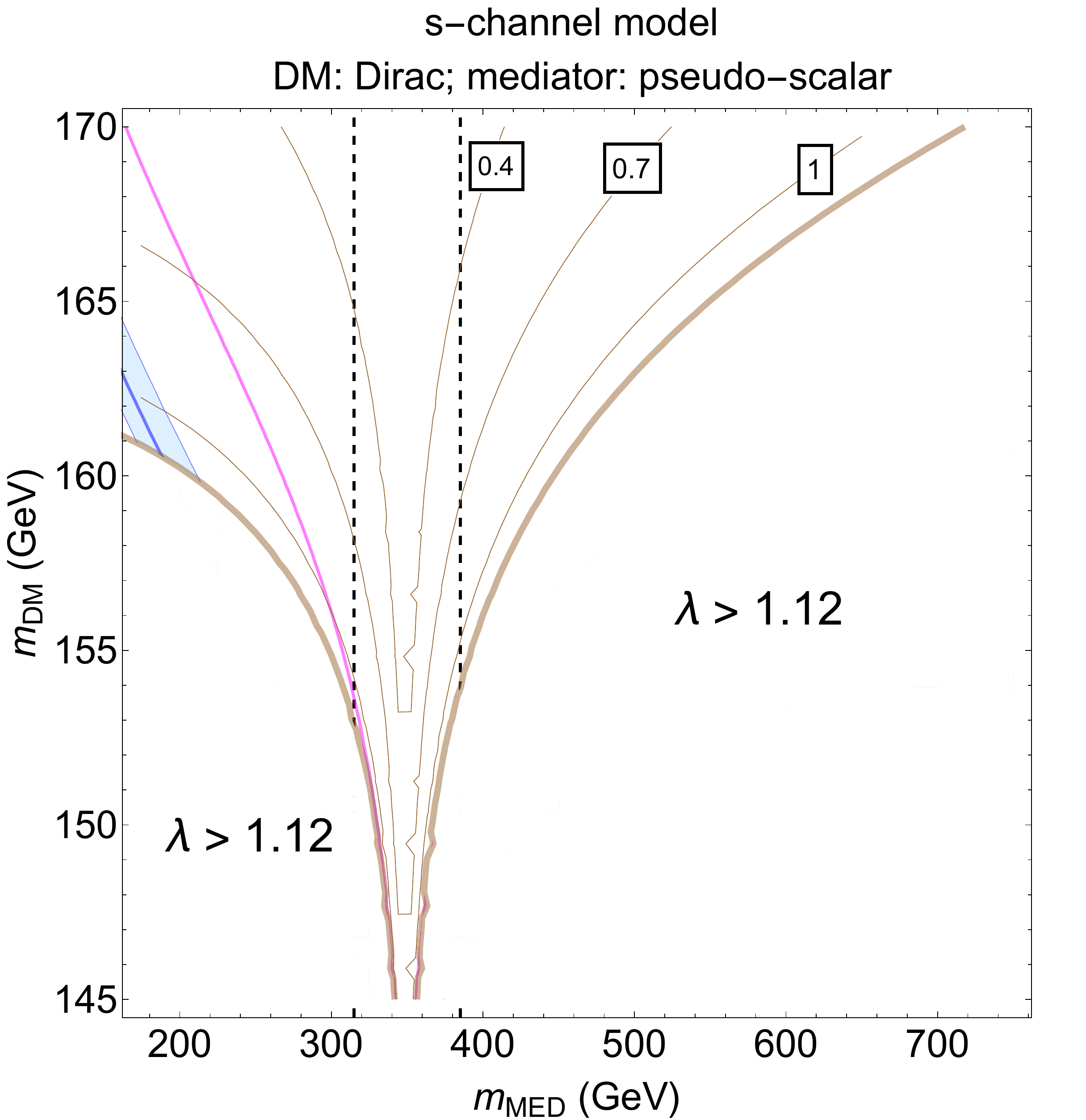}
\caption{
Limits on the $s$-channel model.
The coupling $\lambda$ is fixed here with the observed DM abundance; the brown curves denote these contours of $\lambda$.
The blank region is where $\lambda > 1.12$ and is thus non-perturbative; 
see the text for how this bound is estimated.
The area under the blue curve is excluded at 95\% C.L. by monojet searches at the 8 TeV LHC with $\lag = 20~{\rm fb}^{-1}$. 
The bands around the curve denote uncertainties due to renormalization and factorization scales.
The area under the magenta curve is coverable by the 95\% C.L. projection limits provided by ATLAS for $\sqrt{s}$ = 14 TeV and $\lag = 3000~{\rm fb}^{-1}$.
On account of adopting an EFT treatment, these monojet bounds are more stringent than true limits. 
See the text for more details.
In the region bounded by the dashed curves, the determination of $\lambda$ by Taylor-expanding $\sigmaveeave$ in $v$ may be inaccurate.}
\label{fig:sch}
\end{center}
\end{figure}

\subsection{$s$-channel mediator} 
\label{subsec:sch}

Since we want $\dm$ to annihilate to SM, we must ensure that the channel $\dm \dm \ra \smed \smed$ is closed.
Therefore we impose $\msmed > \mtop$.  
The mediator $\smed$ can generically have both scalar and pseudoscalar couplings to SM fermions, which would explicitly violate CP and confront strict constraints from neutron EDM measurements. 
For this reason, $\smed$ is usually taken to be a CP eigenstate (and all couplings taken real).
If we admit only CP-even couplings, spin-independent direct detection limits nearly exclude the space of our parameters.
The only region spared is in the neighborhood of $\msmed \simeq 2 \mtop$, where resonant DM annihilation allows 
$\lambda$ to be small.
CP-odd couplings, on the other hand, are well-shielded from spin-independent direct detection searches \cite{Freytsis:2010ne,Dienes:2013xya}, leading to more interesting phenomenology.
Consequently, we will take $\smed$ to be a pure pseudoscalar.
This choice restricts the spin of our DM.
If $\chi$ is a scalar, its couplings to $\smed$ would introduce a new source of large CP violation.
Therefore, we take $\chi$ to be a Dirac fermion; the alternative choice of a Majorana fermion would lead to similar results.

We now turn to the constraints on this model.
The strongest limits are placed by collider searches for monojets and missing momentum \cite{Buckley:2014fba,
Haisch:2012kf,
Fox:2012ru,
Lin:2013sca,
Haisch:2013fla,
Haisch:2015ioa,
Arina:2016cqj}.
This signature is generated by the production of $\smed$ and radiation of a gluon, such as in the Feynman diagrams in Fig.~\ref{fig:schFeyns}.
A potentially competitive probe is $t\bar{t} + \met$ final states, but as pointed out in \cite{Haisch:2015ioa}, is always weaker than the monojet search when $\smed$ is CP-odd.
A $t\bar{t}$ resonance via the one-loop process $gg \ra \smed \ra t\bar{t}$ can modify top-pair production by $\mathcal{O}(1\%)$ \cite{Haisch:2013fla}, but is unlikely to be found at the LHC since the theoretical uncertainty on the cross-section is already $\sim 5 \%$~\cite{Czakon:2013goa}.
A dijet resonance via the two-loop process $gg \ra \smed \ra gg$ is also possible, but again unresolvable due to loop suppression~\cite{Haisch:2013fla}.
The imposition of MFV implies that the couplings of $\smed$ to the lighter quarks are suppressed with respect to $\lsm$ by factors of $m_q/\mtop$.
Therefore, they are beyond the sensitivity of LHC searches.
See also Ref.~\cite{Dolan:2016qvg}, which performs a sensitivity study of the mediator in trying to determine its quantum  spin and CP properties.
The introduction of the singlet mediator could have potentially given rise to mixing in the Higgs sector, invoking limits from Higgs coupling measurements at the LHC.
However, by choosing $\smed$ to be CP-odd, mixing with the SM Higgs is pre-empted. 
Finally, no limits from direct detection experiments apply to the region of parameters considered here
since both the spin-dependent and spin-independent cross-sections generated are too small to be probed. 

The limits are sketched in Fig.~\ref{fig:sch} on the $\mdm-\mmed$ plane.
Here $\lambda$ is fixed by requiring $\relabund = 0.1197$.
Contours of $\lambda$ thus obtained are represented by brown curves.
There is a value of $\lambda$ above which the theory becomes non-perturbative, which we estimate as follows. 
As mentioned in Appendix~\ref{app:formulae}, the $\lsm\smed\bar{t}t$ vertex  must arise from a term such as $c H \phi\bar{Q}_3  t^c/\Lambda$ in the unbroken electroweak phase, where $c$ is an $O(1)$ coefficient, $\Lambda$ is a new physics scale and $\phi$ is a complex scalar containing $\smed$.
The coupling $\lsm$ is then identified with $c\cdot v/\Lambda$, where $v$ is the Higgs vacuum expectation value.
For the effective theory to be valid, we require $v/\Lambda$ to be at most $\sim 1/10$.
This fixes the maximum size of $\lsm$, namely 0.1.
No such constraint applies for $\ldm$, which can take on its maximum allowable perturbative size of $4\pi$.
Together, we obtain the upper bound $\lambda \leq \sqrt{(0.1)(4\pi)} \approx 1.12$.
The blank region in Fig.~\ref{fig:sch} is where $\lambda > 1.12$, becoming non-perturbative and invalidating the effective theory.
In the range of paramaters shown, we varied $\decay$ subject to $\lambda = \lsm \ldm$, and find no observable effect on $\relabund$.
(Expressions for $\decay$ in terms of other model parameters can be found in \cite{Haisch:2015ioa,Abdallah:2015ter}.)
Therefore, we do not present constraints on $\decay$. 

The blue curve denotes the 95\% C.L. exclusion from the CMS monojet + $\met$ search at $\shat = 8$ TeV with $\mathcal{L} = 20~{\rm fb}^{-1}$ \cite{CMS8TeVmonojet}.
We derive this curve by recasting the limits obtained in \cite{Haisch:2015ioa}, which had used an EFT description for setting bounds: the DM-top interaction was assumed a contact operator, with the mediator $\smed$ integrated out.
The EFT suppression scale is given by 
\begin{equation*}
\Lambda_{\rm EFT} = \left(\frac{\mtop\msmed^2}{\lambda^2}\right)^{1/3}~.
\end{equation*}
The limit obtained was $\Lambda_{\rm EFT} = 170$ GeV.
The common renormalization and factorization scale in this analysis is $\mu_Q \equiv\left(\sqrt{m^2_{\dm \dm}+p^2_{T,j_1}} + p_{T,j_1}\right)/2$, where $m_{\dm \dm}$ is the $\chi$-$\bar{\chi}$ invariant mass and $p_{T,j_1}$ is the $p_T$ of the hardest jet in an event.
The bands around the blue curve corresponds to the theoretical uncertainty from varying this scale up and down by a factor of 2.
The range of uncertainty thus obtained in the limits is $\Lambda_{\rm EFT} \in [160,185]$ GeV.
These bounds are conservative on account of assuming a contact interaction between the DM and top. 
This is best understood by inspecting the propagator in both the full theory and the EFT treatment.
If the full theory is used, the cross-section is suppressed when the propagator goes off-shell, which may happen for a tight cut on $\met$. 
On the contrary, in the EFT treatment, cutting hard on $\met$ leads to a gross {\em overestimate} of the cross-section.
This happens because a tight cut on $\met$ increases the energy of the final state, or the energy running through the loop, making it approach the suppression scale $\Lambda_{\rm EFT}$. 
The EFT treatment then becomes inaccurate and does not capture the suppression of production rates.
See \cite{Fox:2012ru} and \cite{Haisch:2015ioa} for studies on the differences between an EFT approach and treating the mediator as an active degree of freedom.

The bound in Fig.~\ref{fig:sch} was obtained by imposing $\met > 450~$GeV \cite{Haisch:2015ioa}.
Therefore the propagator would have been sent off-shell in the full theory and the cross-section diminished.
This has had the effect of rendering the EFT approach very crude. 
Consequently, any bound in the region $\msmed < 450$ GeV must be interpreted with extreme caution, and the true limit remembered to be weaker.

We also recast the future sensitivity of the monojet + $\met$ search at $\shat$ =14 TeV and $\mathcal{L} = 3000~{\rm fb}^{-1}$, keeping with the EFT approach.
This limit had been obtained from an ATLAS sensitivity study for this energy and luminosity \cite{ATLAS14TeVprospex}.
We denote it by the magenta curve, which corresponds to $\Lambda_{\rm EFT} = 250$ GeV.
The renormalization and factorization scale here is $2\mu_Q$, and the scale uncertainty was not presented in this case.
We notice from this plot that, even with the most optimistic reach of the LHC, the parameter space of this model is not probed well. 
One concludes that much room is left for the $s$-channel forbidden WIMP scenario. 
One possibly better means to probe the uncovered regions would be a future 100 TeV collider.
Whether monojet production or $t$-$\bar{t}$ production would be the better probe can only be answered with a detailed analysis involving well-chosen cuts. 
Such an analysis is beyond our current scope.
We re-emphasize that, due to the EFT treatment, the reach denoted by the magenta curve is merely an order-of-magnitude estimate,
and must be interpreted as an upper limit on the actual reach.

Finally, a remark on the accuracy of the above limits is in order. 
The Taylor expansion $\sigmaveeave \approx a + b~v^2$ may not be entirely valid close to the pole of the annihilation. 
The discrepancy between naive analytical approximations and a full numerical treatment in computing $\sigmaveeave$ near the pole is described in  \cite{Griest:1990kh} and \cite{Gondolo:1990dk}.
Here ``near the pole" means the region within about 10\% of the pole mass.
We indicate this region with dashed lines in Fig.~\ref{fig:sch}, which encompasses $\msmed \in [310,390]$ GeV.\footnote{The resonant peak in the thermal cross-section with an $s$-channel mediator is usually near twice the DM mass. 
In the case of forbidden DM, it is near twice the mass of the final state. 
This is because in forbidden DM phase space, most of the annihilation comes from DM momenta near the annihilation threshold (which in our case is $2 \mtop$). 
In contrast, usual DM annihilation proceeds mostly near zero DM momentum.}
Our estimate of $\lambda$ in this range is unreliable and smaller than the actual value. 
The discrepancy between the Taylor expansion and the full treatment grows starker with diminishing $\decay/\mmed$ \cite{Griest:1990kh,Gondolo:1990dk}, which can happen in our scenario when $\mdm > 155$ GeV and the decay mode $\smed \ra \dm \dm$ is absent. Fortunately, the area bounded by the 8 TeV monojet search does not fall within this region.

Taking into account the above limits and considerations, we find that the allowed range of parameters covers almost the entire region where the perturbativity bound $\lambda < 1.12$ is imposed. 
Specifically, when perturbative, the model is allowed to reside in the mass ranges $\mdm \in [145, 170]$ GeV and $\msmed \in [175, 725]$ GeV.
As one expects, the allowed range of DM masses is at its largest in the ``funnel region", $\msmed \sim 2 \mtop$.

In the next sub-section we explore the viability and limits of the forbidden mechanism with a $t$-channel mediator.

\begin{figure*}
\begin{center}
\includegraphics[width=.45\textwidth]{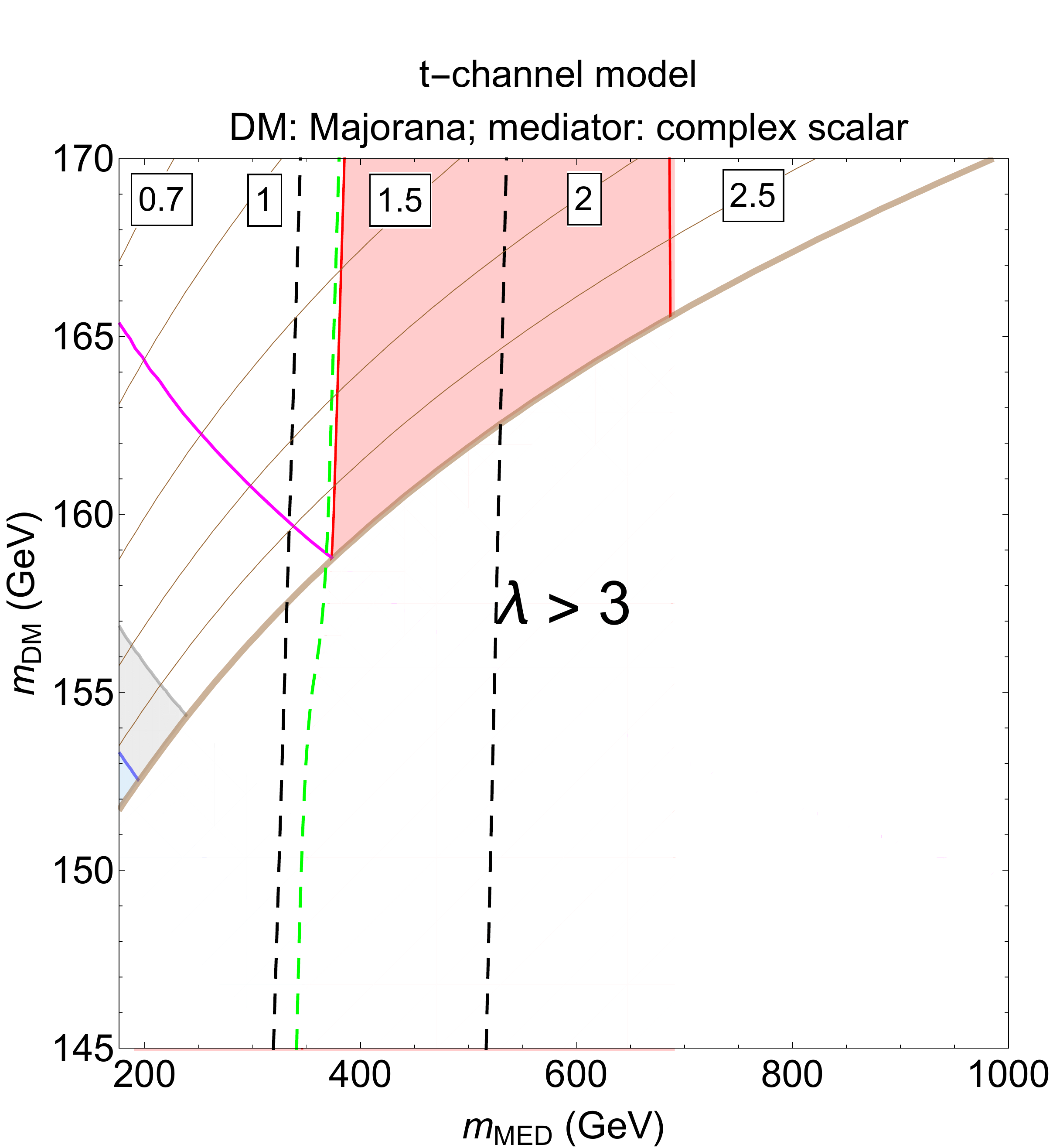}
\quad \quad
\includegraphics[width=.45\textwidth]{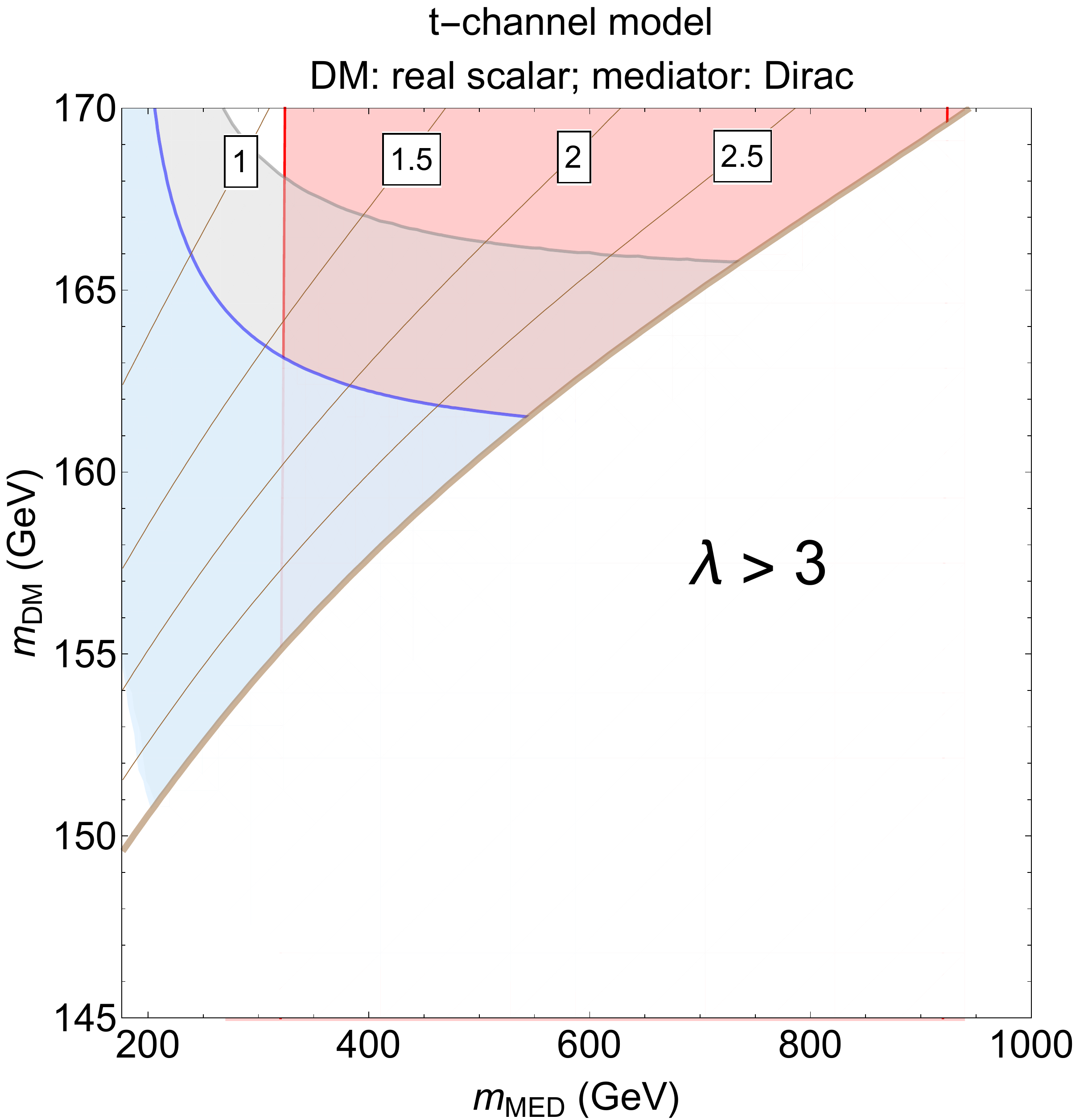}
\caption{
Limits on the $t$-channel models, with Majorana fermion (real scalar) DM on the left (right).
The coupling $\lambda$ is fixed with the observed DM abundance; the brown curves represent contours of this coupling.
The blank region corresponds to $\lambda > 3$. 
The blue shaded regions are excluded at 90\% C.L. by LUX, the gray shaded regions excluded at 90\% C.L. by PandaX-II, and the magenta curve denotes the 90\% C.L. future sensitivity reach of XENON1T.
The red shaded regions are excluded at 95\% C.L. by top squark searches at the LHC; 
the dashed green curve on the left denotes the 5$\sigma$ reach projected by ATLAS at $\sqrt{s} = 14$ TeV and $\lag = 3000~{\rm fb}^{-1}$.
The black dashed curves on the left encompass the 5$\sigma$ reach at $\sqrt{s} = 13$ TeV and $\lag = 3000~{\rm fb}^{-1}$ claimed by the strategy presented in  \cite{An:2015uwa} for the ``compressed region" of stops.
For more details, see the text.}
\label{fig:tch}
\end{center}
\end{figure*}

\subsection{$t$-channel mediator} 
\label{subsec:tch}

Here we consider two models, one with a Majorana fermion DM and a complex scalar mediator,
and another with a real scalar DM and Dirac fermion mediator. 
To avoid confusion over terminology, we label these models ``\tchlabelf" and ``\tchlabels" respectively.
In \tchlabelf, the choice between Majorana and Dirac DM is slight. 
As explained e.g. in \cite{Chang:2013oia}, Majorana DM annihilation picks up a chirality flip in the $s$-wave, so that $\sigmavee_{\rm Maj} \approx (m_f/\mdm)^2 \sigmavee_{\rm Dirac}$, where $m_f$ is the mass of the annihilation products.
Since in our case $m_f = \mtop$ is close to $\mdm$, the results for Majorana and Dirac are found to be qualitatively the same.
In \tchlabels, 
the dominant $s$-wave contribution is four times larger for a real scalar DM than for a complex scalar DM.
Since the annihilation cross-section scales as $\lambda^4$, the couplings are correspondingly only $4^{1/4} \simeq 1.4$ times smaller in the former case.
Therefore no qualitative difference exists between the use of a real and complex scalar for DM.

The constraints on these models come from collider and direct DM detection experiments.
Searches for top superpartners (stops) can be directly recast to our scenario.  
The relevant search here is for pair-production of stops followed by decay to a pair of top quarks + $\met$.
This signature applies to our model since at the LHC our mediator, like the stop, is QCD-produced and decays to top quark + DM.
The case of the scalar mediator of \tchlabelf ~is particularly interesting. 
In both \tchlabelf ~and the MSSM, the top-flavored scalar is produced entirely through QCD and the BR = 100\%.\footnote{This is not the case had other mediator flavors been present, e.g., the production of the first two generations of mediators would involve processes with $t$-channel DM exchange, at variance with similar processes of squark production because of the difference in couplings.}
Hence, due to identicality of production and decay, the exclusion region for the scalar mediator of \tchlabelf ~is congruent to that of the top superpartner (for LSP masses that range in our DM masses).
However, the rate of the fermion mediator production through QCD can be different in \tchlabels, as we explain below.
Therefore to set bounds, we first generate the mediator pair production process $pp \ra \med \bar{\med}$ in {\tt MadGraph5} \cite{Alwall:2014hca} and obtain the production cross-sections at leading order. 
Then we apply a $K$-factor of 1.5 (see \cite{Anandakrishnan:2015yfa}) to obtain the NLO cross-section.
Assuming similar acceptances for the production of a stop pair and $\med \bar{\med}$, we compare our rates with the 95\% C.L. exclusion cross-sections provided in \cite{CMS:2014wsa} and find the limits on $\mmed$.

DM in \tchlabels, \tchlabelf\, can scatter with nuclei through gluon loops 
(see \cite{Hisano:2015bma} for all the attendant Feynman diagrams), 
contributing to spin-independent direct detection rates.
These limits are shown on the $\mdm-\mmed$ plane in Fig.~\ref{fig:tch}, where the plot on the left (right) corresponds to \tchlabelf~(\tchlabels).
As in our treatment of the $s$-channel model, we fix $\lambda$ throughout these plots by the requirement $\relabund = 0.1197$ and denote these contours of $\lambda$ with brown curves. 
The blank region in both plots is where $\lambda \geq 3$, violating perturbativity of the coupling.\footnote{Here we obtain this value from a conservative perturbativity condition: $\lambda^2/(8\pi^2) \lsim 0.1$.
Furthermore, for $\lambda > 3$, its rapid RG running produces a Landau pole below a scale of 10 TeV, that we wish to avoid.}
In the plot we require $\mmed > \mdm$ since we require $\dm$ to be the lightest field charged odd under $Z_2$.
While nothing prevents $\mdm < \mmed < \mtop$, we choose to present limits in the region $\mmed > \mtop$.
This is the range relevant to LHC searches in the $\med \ra \dm t$ channel; in addition, the effects of co-annihilation between $\med$ and $\dm$ are mitigated by separating $\mmed$ and $\mdm$.
Although co-annihilations are in principle allowed in our models, we wish to focus solely on the effects of the forbidden mechanism.

We now describe the various curves in Fig.~\ref{fig:tch}. The red shaded region in both plots is ruled out at the 95\% C.L. limit by ATLAS stop searches at $\shat = 8$~TeV with $\lag = 20~{\rm fb}^{-1}$  \cite{Aad:2015pfx}.
CMS sets similar bounds \cite{Khachatryan:2015wza,CMS:2014wsa}.
We now show that the future prospects of these models are optimistic -- most of the parameter space in the perturbative region can be probed in upcoming LHC searches.
The region to the right of the dashed green curve in the left-hand plot of Fig.~\ref{fig:tch} denotes the $5\sigma$ discovery reach at $\shat = 14$~TeV and $\lag = 3000~{\rm fb}^{-1}$ as projected by ATLAS in \cite{ATLASfuture}.
This reach is provided in the projected search combining 0-lepton and 1-lepton channels.
The other end of this reach is at a mass of $\mmed$ = 1400~GeV, well outside our perturbative region.
Projections in these channels have also been made by CMS \cite{CMSfuture} but at a lower luminosity of $\lag = 300~{\rm fb}^{-1}$, due to which the reach does not appear as optimistic as the prospects presented by ATLAS.
Since no analogous projection studies have been provided for the fermionic mediator of \tchlabels, we do not provide their sensitivities.

 Another interesting prospect for these models is to hunt for the mediator in the ``compressed" region near $\mmed \approx \mdm + \mtop$. 
A strategy dedicated to compressed stop searches was introduced in \cite{Hagiwara:2013tva,An:2015uwa}, exploiting the recoil of ISR jets against the $\met$ seen in these parametric regions.
In \cite{An:2015uwa} the projected reach for a 5$\sigma$ discovery by this strategy for top squarks at $\shat = 13~$TeV and $\lag = 3~{\rm ab}^{-1}$ was presented.
We show this on the left-hand plot in the region flanked by the dashed black curves.
The technique can be seen to cover the compressed region all the way down to $\mmed = \mdm +\mtop$. 
Analogous sensitivities for \tchlabels~were not presented, however we may reasonably expect the same technique to cover the entire compressed region again.

Finally, the blue and gray shaded regions in the two plots are ruled out at the 90\% C.L. limit by spin-independent scattering cross-section bounds set by LUX \cite{Akerib:2013tjd} and PandaX-II \cite{Tan:2016zwf} respectively. 

The magenta curve on the left-hand plot depicts the future sensitivity (at 90\% C.L. limit ) at the XENON1T experiment \cite{Aprile:xenon1T} proposed to go live in the year 2017.
On the right-hand plot, the corresponding magenta curve would fall outside the range of masses shown.
In the region left uncovered by PandaX-II, the smallest spin-independent scattering cross-section is $\sigma_{\rm SI}^{\rm scal} \simeq 2 \times 10^{-45}~{\rm cm}^2$.
This is about two orders of magnitude larger than the reach of XENON1T, and can thus be comfortably covered by the experiment.

When we compare across the two plots, we notice a difference: 
 while only a small region in the perturbative region of \tchlabelf ~is seen to be excluded by LUX, a much wider range (upto $\mmed \simeq$ 700 GeV) is covered for \tchlabels.
One understands this difference from the interplay between the annihilation and direct detection cross-sections, as follows. 
By approximating $\mmed \gg \mtop$, the annihilation cross-section for both fermionic and scalar DM can be parametrized as $\sigma_{\rm ann} \sim \lambda^4 \mdm^2/\mmed^4$.
In the DM-nucleon scattering cross-section, on the other hand, there are two key differences between \tchlabelf ~and \tchlabels. 
First, consider the DM-DM-gluon-gluon effective operators in the two models:

\bea
\nn f_N \frac{\alpha_s}{\pi}\bar{\dm}\dm \mathcal{G}_{\mu \nu}\mathcal{G}^{\mu \nu},~& &{\rm for~\tchlabelf,~and}  \\
f_N \frac{\alpha_s}{\pi}\dm^2 \mathcal{G}_{\mu \nu}\mathcal{G}^{\mu \nu},~& &{\rm for~\tchlabels}.
\label{eq:ggDMDMops}
\eea


The Wilson co-efficients $f_N$ scale as
\bea
\nn \mdm/\mmed^4,~& &{\rm for~\tchlabelf,~and}  \\
\nn 1/\mmed^2,~& &{\rm for~\tchlabels}.
\eea
The difference in the mass dimensions of the $f_N$'s reflects the fact that fermionic and scalar DM fields have different dimensions.
More crucially, a factor of $\mdm$ appears in the numerator of the fermion $f_N$. 
This factor can be understood by the fact that loop level effects (such as Eq.~(\ref{eq:ggDMDMops})) generated from a renomalizable theory must respect the same symmetries as the underlying theory. 
Here, the operator in the \tchlabelf\, case violates chiral symmetry and therefore must vanish as $\mdm \to 0$ and chiral symmetry in the underlying theory is restored. 
Second, the DM-nucleon scattering cross-section of scalar DM is suppressed in the denominator by a factor of $\mdm^2$, which is absent for fermion DM.
This extra scaling arises from the ratio of the matrix elements
\begin{equation*}
\frac{\langle \chi| \chi^2 |\chi \rangle_{\rm scalar}}{\langle \chi | \bar{\chi} \chi | \chi \rangle_{\rm fermion}} \propto \frac{1}{\mdm}~. 
 \end{equation*}
The resulting spin-independent direct detection cross-sections go as 
\bea
\nn \sigma_{\rm SI}^{\rm ferm} &\sim& \frac{\lambda^4 m_N^4\mdm^2}{\mmed^8}; \\
\nn \sigma_{\rm SI}^{\rm scal} &\sim& \frac{\lambda^4 m_N^4}{\mdm^2\mmed^4},
   \eea
 where $m_N$ is the nucleon mass.
These can be written as 
\beq
\sigma_{\rm SI}^{\rm ferm} \sim \sigma_{\rm ann}m_N^4/\mmed^4;~~ \sigma_{\rm SI}^{\rm scal} \sim \sigma_{\rm ann} m_N^4/\mdm^4~.
\label{eq:DDannlinks}
\eeq
Since we fix the relic abundance to the observed value throughout our plots, we can now write a direct comparison between the fermion and scalar DM scattering cross-sections for a given combination of DM and mediator mass:
$\sigma_{\rm SI}^{\rm ferm} \sim \sigma_{\rm SI}^{\rm scal} (\mdm/\mmed)^4$.
Hence \tchlabelf ~direct detection rates are relatively suppressed, giving us weaker bounds and less optimistic projections.
Eq.~(\ref{eq:DDannlinks}) also helps one understand the shape of the curves in Fig.~\ref{fig:tch}.
We find that the \tchlabelf ~bound falls quicker than \tchlabels ~with respect to $\mmed$.
This is due to the $1/\mmed^4$ scaling of $\sigma_{\rm SI}^{\rm ferm}$ that is not seen in $\sigma_{\rm SI}^{\rm scal}$.
At $\mmed \gg \mtop$, we find the \tchlabels ~bound to be insensitive to $\mmed$.
This is because for a fixed $\sigma_{\rm ann}$, $\sigma_{\rm SI}^{\rm scal}$ does not scale with $\mmed$.
}

The collider limits are stronger for fermion mediator pair production in \tchlabels.  
This is because (a) fermions have more spin degrees of freedom than scalars, (b) scalar mediator production ({\em\`a la} stop production) in \tchlabelf~is suppressed by a momentum-dependent coupling in production modes where a gluon mediates in the $s$-channel and $\med$ mediates in the $t$-channel; this is absent in the corresponding fermion mediator production process in \tchlabels.
The larger production rates of the fermion mediator results in two consequences: at the upper end of the mass bound, the reach is higher, while at the lower end, near the compressed region $\mmed \simeq \mdm$, the fermion mediator is able to better overcome the dwindling signal acceptance of the search.
Hence we find that while $\mmed \in [380,690]$~GeV is excluded in \tchlabelf, the corresponding range is $\mmed \in [310,920]$~GeV in \tchlabels. 
The latter is in good agreement with the recasting performed by the authors of \cite{Anandakrishnan:2015yfa} for fermionic top partners that are odd under a parity.

The Majorana DM model has a supersymmetric limit, where $\dm$ is a neutralino LSP of the MSSM, $\med$ is a top squark.  
In this limit, the coupling $\lambda$ is of electroweak strength and hence feeble.
Therefore, it may naively preclude forbidden annihilation of neutralino DM as a viable option, but this can be circumvented with co-annihilation effects.
More discussion on co-annihilation and supersymmetry is relegated to the final section of the paper, Sec.~\ref{sec:discs}.

In Sec.~\ref{sec:simpmods} we mentioned the possibility of a ``bottom portal" forbidden WIMP with a mediator evading LEP bounds.
Such a theory can be built in a manner analogous to the models seen in this sub-section, with $t_R \ra b_R$ and the $t$-channel mediator $\med \ra \widetilde{B}$. 
One then asks what couplings and DM masses are possible in this scenario.
The heavier $\widetilde{B}$ is, the more inefficient is the annihilation, and hence the stronger is the coupling and/or the more degenerate is $\mdm$ with $m_{\rm bottom}$ = 4.2 GeV, if we need to achieve the correct DM abundance.
To maximize the forbidden annihilation cross-section, let us set the mediator mass $m_{\widetilde{B}} = 104$~GeV, which is the edge of the LEP limit~\cite{LEPcino,LEPslep}, and the coupling $\lambda = 3$, its maximum perturbative size.
If we now ask what value of $\mdm$ yields the correct DM abundance, we find 
\bea
\nn \mdm =& 4.0~{\rm GeV},\  &{\rm for\ Majorana\ fermion\ DM}, \\
\nn \mdm =&~3.95~{\rm GeV},\  &{\rm for\ real\ scalar\ DM}.
\eea
The above DM masses are the minimum required, since larger $m_{\widetilde{B}}$ and/or smaller $\lambda$ values
reduce $\sigmaveeave$, compelling $\mdm$ to move closer to $m_{\rm bottom}$ to overcome the Boltzmann suppression.
Thus the range of parameters where this scenario is viable is very limited, and we do not pursue it further.

\subsection{Indirect detection limits} 
\label{subsec:indirect}

Constraints on the present-day annihilation cross-section of DM are set by Fermi-LAT \cite{Ackermann:2013yva} by the observation of 15 dwarf galaxies.
For a thermal cross-section $\sigmaveeave = 3 \times 10^{-26}~{\rm cm}^3~{\rm s}^{-1}$, the 95\% C.L.  limit is $\lsim$ 95 GeV in the $b$-$\bar{b}$ final state, which is the strongest constraint.
Our $\mdm \supset [145, 170]$~GeV is above this limit. 
Moreover the present-day forbidden annihilation to top quark pairs is exponentially suppressed as $\exp(-2(\mtop-\mdm)/T_{\rm today})$, where $T_{\rm today} = 2.7$~K.
This cross-section is several orders of magnitude below $3 \times 10^{-26}~{\rm cm}^3 ~{\rm s}^{-1}$.

There are kinematically ``allowed" final states in our models from loops, which include $\gamma, g, Z$ and $h$.
Ref.~\cite{Jackson:2013rqp} considers these annihilation channels; their cross-sections are enhanced by assuming an $s$-channel vector mediator near resonance, coupled strongly to a vector-like fermion that mixes with the top quark.
In regions where the DM mass is close to the top quark's, the relic density was not calculated accounting for the forbidden mechanism.
This was because the focus of \cite{Jackson:2013rqp} was to demonstrate the possibility of non-trivial indirect detection signals in a forbidden set-up.
However, since our focus is on the relic density, these ``allowed" channels are always safe from indirect detection bounds. 
One sees this by considering that our DM freezeout was dictated by forbidden annihilation, in comparison to which the light final state channels are sub-dominant due to loop-suppression.
This automatically insures that the latter have a thermal cross-section much smaller than $3 \times 10^{-26}~{\rm cm}^3 ~{\rm s}^{-1}$, and will therefore never saturate indirect detection bounds.
In a similar manner, the $b$-$\bar{b}$ final state from loops, as well as three-body annihilations to such states as $tWb$, are also expected to be unconstrained by line and diffuse photon searches.

\begin{figure}
\begin{center}
\includegraphics[width=.45\textwidth]{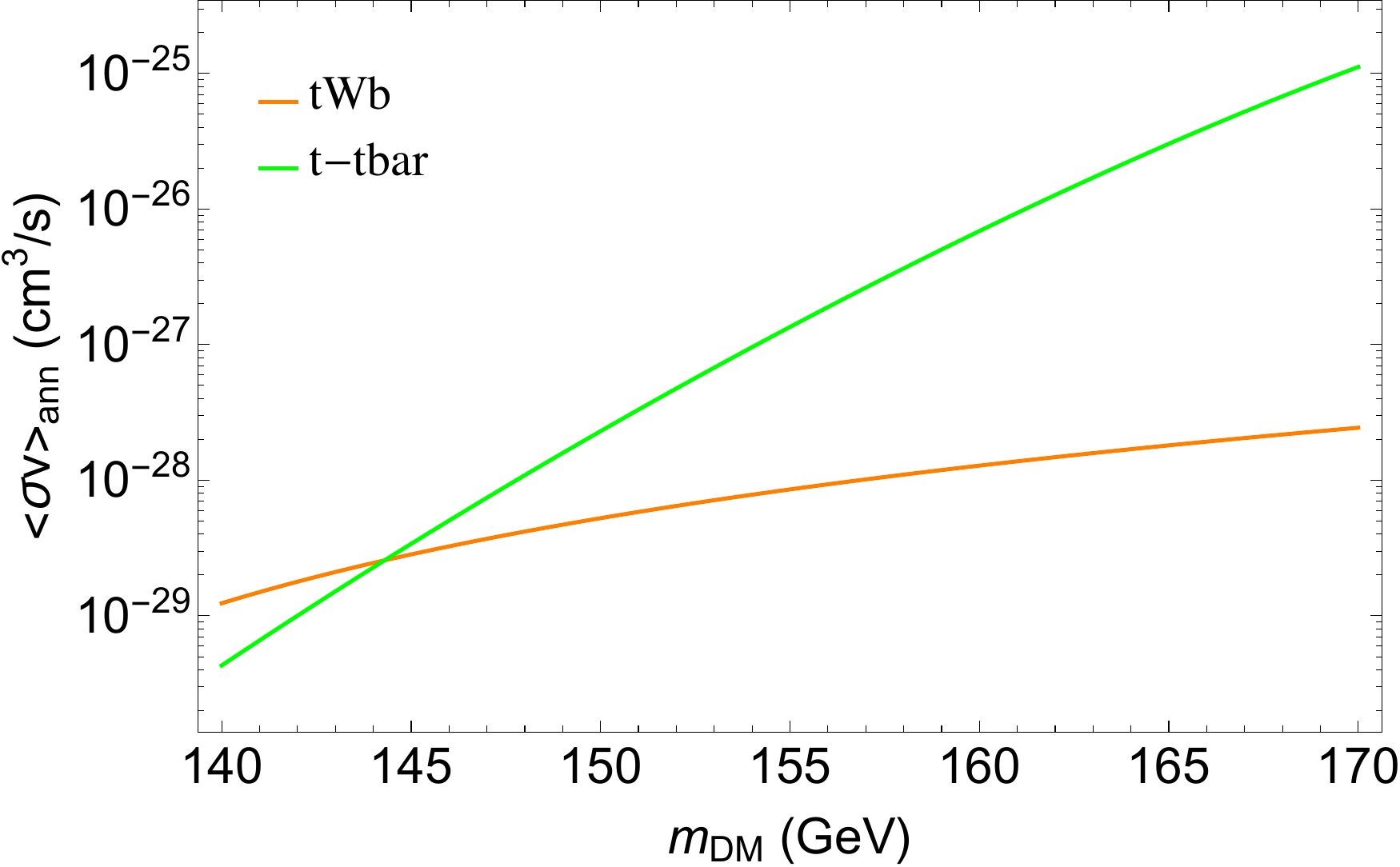}
\caption{
A comparison of the thermal cross-sections of 2-body ($\dm \dm \ra t \bar{t}$)  and the dominant 3-body ($\dm \dm \ra t W b$) final states in our forbidden WIMP set-up.
For illustration, we have chosen the $t$-channel model with fermionic DM and $\lambda = 1$, $\mmed = 180$ GeV. 
The results are similar for other choices of models and parameters.
In spite of the Boltzmann exponential suppression of forbidden annihilation, the 2-body process dominates over the phase space-suppressed 3-body annihilation for $\mdm > 145$ GeV.
This range happens to enclose the region where our $\lambda$ is perturbative (Figs.~\ref{fig:sch},~\ref{fig:tch}).}
\label{fig:2v3}
\end{center}
\end{figure}

\subsection{Three-body final states} 
\label{subsec:3body}

In presenting the limits on our models above, we assumed that the thermal cross-section of the $t\bar{t}$ final state overwhelms that of a three-body final state $tWd_i$, where $d_i$ is a down-sector quark.
It is worthwhile to check the accuracy of this assumption.
It was found in \cite{Jackson:2013rqp,Chen:1998dp,Yaguna:2010hn} that for DM masses $\mdm \lsim \mtop$, the $\dm\dm \ra tWb$ process can have rates near thermal cross-sections for certain model parameters.
In these models the forbidden annihilation to $t\bar{t}$ was neglected.
In our case, at large $\mtop$-$\mdm$ splittings it is possible that the Boltzmann exponential suppression of annihilation to $t\bar{t}$ compares to the three-body phase suppression of annihilation to $tWb$.  
Using analytic expressions provided in \cite{Chen:1998dp}, and setting $\lambda=1$ and $\mmed = 180$ GeV, we compute $\sigmaveeave$ for $tWb$ production and plot it as a function of $\mdm$ in Fig.~\ref{fig:2v3} (the orange curve). 
For comparison we also plot the $t\bar{t}$ thermal cross-section in green with the same parameters.
We have chosen the $t$-channel model \tchlabelf~ for illustration.
For $\mdm > 145$~GeV, the two-body cross-section dominates.
For $\mdm \geq 150$ GeV, the three-body rate is about 10\% or less, and has negligible impact on our phenomenology.
We have checked that this is true for all combinations of $\lambda$, $\mmed$ and models considered here.

Let us now inspect the implications of this finding for our phenomenology.
We have illustrated before that $\mdm > 145$ GeV encloses the region of our parameters where $\lambda$ is perturbative.
It follows that any significant modification the three-body annihilations may have on our experimental limits would be in the mass range $\mdm \in [145, 150]$ GeV.
Since an extra annihilation channel is added, its effect is to decrease the $\lambda$ required for obtaining the right abundance, hence leading to weaker constraints.
In Fig.~\ref{fig:sch}, the region in the range $145~{\rm GeV} \leq \mdm \leq 150~{\rm GeV}$ falls within the dashed curves, where our estimation of $\lambda$ is already unreliable due to the Taylor expansion.
In the plots of Fig.~\ref{fig:tch}, all the area in the region $145~{\rm GeV} \leq \mdm \leq 150~{\rm GeV}$ is already in the non-perturbative region.
These observations indicate that in the region where our constraints apply, three-body final states may be neglected.

\section{Discussion}
\label{sec:discs}

In our work, after having gone through every SM state, we settled on the top quark as the only remaining suitor for a final state that contributes (almost) 100\% to DM annihilation via a forbidden mechanism at the weak scale and couples at the renormalizable level to the DM particle.
And by demanding that DM annihilation be predominantly forbidden, we have obtained the tightest 
constraints possible for a forbidden WIMP.
Even before imposing experimental limits, we find the bound from requiring coupling constant perturbativity squeezing our parameter space into a tight region. 
Specifically, in the $s$-channel model, we are confined to DM masses between 145 GeV and 170 GeV, and mediator masses between 175 GeV and 725 GeV, with the least restrictive region near the ``funnel" at $\msmed \sim 2 \mtop$; in the $t$-channel models, our DM masses could range between 150 GeV and 170 GeV, with the mediator masses between 175 GeV and 1000 GeV.
Applying experimental limits, we find that (i) the $s$-channel model is mostly unconstrained by monojet searches at the LHC (see Fig.~\ref{fig:sch}), 
(ii) the $t$-channel models are constrained to different extents depending on the spin of dark matter. 
The model with fermionic DM (\tchlabelf) is poorly bounded by DM direct detection experiments, but excluded in the mediator mass range [380, 690] GeV by ATLAS stop searches.
The model with scalar DM (\tchlabels) is highly constrained by PandaX-II and the ATLAS search -- these probe the parameter space in complementary regions and collectively exclude most of it (see Fig.~\ref{fig:tch}).

Our bounds, however, may be relaxed if we also allowed for more annihilations to SM states lighter than $\dm$, which would diminish $\lambda$ and reflect in weaker limits from collider and direct detection searches.
There are several ways to arrange this.
For instance, one may allow for a more general coupling structure 
that will allow $\dm$ to annihilate to lighter quarks of both up and down types, and to leptons.
In the $t$-channel models, one may still wish to keep annihilations restricted to the third generation
in order to safeguard against direct detection limits, but allow tree-level $b$-$\bar{b}$ final states
by charging the mediator $\med$ under $SU(2)_W$.
Forbidden final states involving weak boson and Higgs final states may be accommodated if $\dm$ also partially annihilated to light fermions.
All these alterations would expose these models to experimental probes not discussed here; we leave these avenues of study for future work. 

The limits given here may also be weakened if one took into account the effects of co-annihilation.
Were $\mmed$ close to $\mdm$ in the $t$-channel models, the process $\med \dm \ra {\rm gluon}+ {\rm top}$ would dominate the effective thermal cross-section.
New states that assist co-annihilation may also be present.
We may see this illustratively in a supersymmetric context, our analogue of which is the model \tchlabelf.
Consider a bino LSP will all other electroweak-inos decoupled.
If one had a right-handed top squark $\widetilde{t}_R$ not much above in mass, and decoupled all other superpartners, the bino would annihilate predominantly to top quark pairs.
If its mass were just below $\mtop$, only forbidden annihilation is possible.
Given the bino couples with hypercharge, this is an extremely inefficient process and would result in an unacceptably large relic abundance.
But if we imagine in the spectrum the presence of a gluino that is less than $\sim 10\%$ above the bino in mass, co-annihilation may play a crucial role.
The efficient self-annihilation of the gluino would now depopulate the bino abundance, and may set the effective thermal cross-section\footnote{As another example, if we take an LSP that has wino and/or higgsino content, the co-annihilation of the neutral components of the $SU(2)_W$ multiplets with their corresponding charged components may be balanced against the forbidden annihilation of the LSP to heavier SM states.}.
In this way, forbidden annihilation may be incorporated into supersymmetry models that provide DM candidates.
Here one might wonder if a gluino with a mass near $\mtop$ is already ruled out by hadron collider experiments.
However, the ``compressed" nature of the gluino-bino spectrum, begetting soft jets and subdued MET, is unfriendly to the existing jets+$\met$-based searches.
See e.g., Ref.~\cite{Alwall:2008ve}, where various cuts
were optimized to recast Tevatron searches for a simplified gluino-bino spectrum -- one finds the region of bino mass $\lsim \mtop$ uncovered.
To date, no dedicated search has been performed for the spectrum of interest.
 

An intriguing possibility (in both supersymmetric and non-supersymmetric contexts) is the forbidden annihilation of DM through an $s$-channel mediator tempered by co-annihilation with nearby states.
If the mediator mass were close to twice the DM mass, all three ``exceptions" outlined in \cite{Griest:1990kh} may be in action.
That Nature may have turned exceptions into a collective rule is an amusing scenario warranting further study.

It is worth emphasizing a crucial phenomenological difference between the forbidden scenarios considered here (annihilation to top quarks) and that of Ref.~\cite{D'Agnolo:2015koa}  (annihilation to dark photons).
Forbidden WIMPs are only very weakly limited by DM indirect detection due to suppressed annihilation rates in the present day, whereas the requirement of sizeable couplings for forbidden annihilation makes them amenable to collider and direct detection probes.
On the other hand, indirect detection constraints play a decisive role in \cite{D'Agnolo:2015koa} -- when gauge kinetic mixing is introduced, annihilations to light SM states in the present day universe are possible.
At the same time, limits from direct detection and collider physics are weak or non-existent.
This is so because this model is an example of a secluded WIMP \cite{Pospelov:2007mp}. 
The key concept is that direct DM-SM couplings are suppressed but the correct thermal cross-section is obtained by taking mediators light; therefore collider and direct detections limits are significantly weakened, whereas indirect detection signals become important.
In this way, forbidden WIMPs and secluded WIMPs are entirely contrasting ideas.
This is not to say that forbidden WIMPs may never yield signals from the sky. 
Interesting prospects in indirect detection can be raised in our models if DM annihilates to new states that mix with the SM.
As \cite{Pospelov:2007mp} demonstrates, one may have annihilations to a $U(1)'$ boson (as in \cite{D'Agnolo:2015koa}), or to a spin-0 mediator mixing with the Higgs boson (possible in our $s$-channel model if $\smed$ is CP-even) or even to a right-handed neutrino.
In such cases, one may also loosen the imposition of flavor structures discussed in this work, allowing more freedom in the way the mediators couple to multiple SM fields.

If the dark matter abundance is indeed set -- in full or part -- by forbidden annihilation to SM states, it might be argued that the DM mass appears to be tuned to fall just short of the SM mass.
Such spectral tuning is inevitable in forbidden WIMP models, and is somewhat akin to the mass degeneracy seen in scenarios involving co-annihilation, or to the careful arrangement of mass parameters that occasions resonant annihilation.
Our spectrum is perhaps the low energy manifestation of a flavor structure in the ultraviolet that is common to the dark and SM sectors. 
In the $t$-channel models, the large sizes of both the effective couplings and the top Yukawas suggest the presence of some strong dynamics being responsible for the near-degeneracy of the $\chi$-$t$ system.
These hints encourage interesting model-building possibilities.

To conclude, forbidden annihilation is a freezeout scenario that we hope will garner more attention than priorly, from both theoretical and experimental communities wishing to constrain non-standard DM models. 
We have shown that, even at the weak scale, such a mechanism may be principally responsible for making DM as abundant as observed. 
This was arranged with minimal introduction of fields and parameters.
If this mechanism were in action at the weak scale, current collider and direct detection experiments are perfectly poised to probe it.
We hope to have pointed the direction to a significant stone to turn in the WIMP paradigm.

\section*{Acknowledgments}
\label{sec:ack}

Carlos Alvarado,
Joseph Bramante,
Spencer Chang,
Fatemeh Elahi
and
Roni Harnik
have variously improved this work with enlightening conversation.
This work was partially supported by the National
Science Foundation under Grants 
No. PHY-1417118 and No. PHY-1520966.

\appendix

\section{Relic abundance calculation}
\label{app:relabundcalc}

In computing the relic density of forbidden WIMPs we employ the prescription briefly laid out in \cite{D'Agnolo:2015koa}.
In what follows we give a more explicit computation of $\sigmaveeave$ 
using ideas from \cite{Gondolo:1990dk}.

Consider the process $\chi_1 \chi_2 \ra \psi_1 \psi_2$, where $\chi_i (\psi_i)$ are DM (SM) states.
Denoting by $f_k$ the phase space density of a species $k$,  
the principle of detailed balance allows us to write
\beq
f^{\rm eq}_{\chi_1} f^{\rm eq}_{\chi_2} = f^{\rm eq}_{\psi_1} f^{\rm eq}_{\psi_2}~,
\label{eq:detbal1}
\eeq
where $f^{\rm eq}_k$ is the corresponding $f_k$ at equilibrium (thermal and chemical).
Then,
\bea
\nn &&\sum_{\rm spins} \int d\widetilde{P}~ \deltild~ |\mathcal{M}_{\chi_1 \chi_2 \ra \psi_1 \psi_2}|^2  f^{\rm eq}_{\chi_1} f^{\rm eq}_{\chi_2}  \\
&=&\sum_{\rm spins} \int d\widetilde{P}~ \deltild~ |\mathcal{M}_{\chi_1 \chi_2 \ra \psi_1 \psi_2}|^2  f^{\rm eq}_{\psi_1} f^{\rm eq}_{\psi_2}~,
\label{eq:detbal2}
\eea
where $d\widetilde{P} \equiv \ptild_{\chi_1}\ptild_{\chi_2}\ptild_{\psi_1}\ptild_{\psi_2}$ with $\ptild_k \equiv d^3p_k/((2\pi)^32E_k)$, $\deltild \equiv (2\pi)^4 \delta^4(p_{\chi_1}+p_{\chi_2}-p_{\psi_1}-p_{\psi_2})$ and $\mathcal{M}$ is the process amplitude.
The principle of unitarity now gives us
\bea
\nn && \sum_{\rm spins} \int \ptild_{\psi_1}\ptild_{\psi_2}~ \deltild~ |\mathcal{M}_{\chi_1 \chi_2 \ra \psi_1 \psi_2}|^2  \\
&=&\sum_{\rm spins} \int \ptild_{\psi_1}\ptild_{\psi_2}~ \deltild~ |\mathcal{M}_{\psi_1 \psi_2 \ra \chi_1 \chi_2}|^2~.
\label{eq:unitarity}
\eea
Moreover, the particle number densities are given by 
\beq
n_k = \int g_k~f_k~\frac{d^3p_k}{(2\pi)^3}~, 
\label{eq:numden}
\eeq
where $g_k$ is the number of internal degrees of freedom.

Notice that the LHS of Eq.~(\ref{eq:unitarity}) is the quantity
\begin{equation*}
4~g_{\chi_1} g_{\chi_2}~F~\sigma_{\chi_1 \chi_2 \ra \psi_1 \psi_2}~,
\end{equation*}
where
\begin{equation*}
F \equiv \sqrt{(p_{\chi_1}\cdot p_{\chi_2})^2 - (m_{\chi_1}m_{\chi_2})^2}~.
\end{equation*}

The M\o ller velocity is given by
\begin{equation*}
v \equiv (|\mathbf{v}_{\dm_1} - \mathbf{v}_{\dm_2} |^2 - |\mathbf{v}_{\dm_1} \times \mathbf{v}_{\dm_2}|^2)^{1/2}  = F/(E_{\dm_1}E_{\dm_2})~.
 \end{equation*}
For our purposes the M\o ller velocity is nothing but the relative velocity between the incident particles.

With the information above one now obtains a simple relation between the process of interest and its reverse
by inserting Eqs.~(\ref{eq:unitarity}) and (\ref{eq:numden}) into Eq.~(\ref{eq:detbal2}):
\beq
\sigmavee_{\chi_1 \chi_2 \ra \psi_1 \psi_2} =  \frac{n^{\rm eq}_{\psi_1}n^{\rm eq}_{\psi_2}}{n^{\rm eq}_{\chi_1}n^{\rm eq}_{\chi_2}} \sigmavee_{\psi_1 \psi_2 \ra \chi_1 \chi_2}~. 
\label{eq:thermalXSmaster}
\eeq
Applying this to $\dm \dm \ra t \bar{t}$, we have 
\beq
\sigmavee_{\dm \dm \ra t \bar{t}} =  \left(\frac{n^{\rm eq}_t}{n^{\rm eq}_\dm}\right)^2 \sigmavee_{t \bar{t} \ra \dm \dm}~. 
\label{eq:thermalXSspecific}
\eeq
The equilibrium number density for a non-relativistic particle of mass $m_k$ is $g_k (m_kT/2\pi)^{3/2}\exp(-m_k/T)$. 
Defining the fractional mass deficit $\delta \equiv (\mtop - \mdm)/\mdm$ and $x\equiv\mdm/T$, we have 
\beq
\sigmavee_{\dm \dm \ra t \bar{t}} =  \left(\frac{g_t}{g_\chi}\right)^2 \sigmavee_{t \bar{t} \ra \dm \dm}(1+\delta)^3 e^{-2\delta x}~. 
\label{eq:thermalXSexpsup}
\eeq
 Accounting for spin and color degeneracies, the degrees of freedom number as $g_t = 2 \times 3$ and $g_\chi = 1~(2)$ for scalar (fermion) DM. Thus the forbidden cross-section is exponentially suppressed, as one would expect from averaging over the tail of the Maxwell-Boltzmann distribution.
This result is consistent with the original prescription of Griest and Seckel \cite{Griest:1990kh}, where the thermal averaging was performed by integration of the DM annihilation cross-section over the range of DM velocities where annihilations occur\footnote{It was claimed in \cite{D'Agnolo:2015koa} that the result of \cite{Griest:1990kh} had an incorrect factor of 2, but we found no such discrepancy.}.

Eq.~(\ref{eq:thermalXSexpsup}) can now be used to solve for the freezeout condition by the usual approximations  \cite{Gondolo:1990dk}.
One assumes here that the top quarks produced by DM annihilation are in equilibrium with the thermal bath, which is reasonable since the top is electrically charged.
The task then simplifies to computing $\sigmavee_{t \bar{t} \ra \dm \dm}$, Taylor-expanded as $a + b~v^2$.
Appendix~\ref{app:formulae} provides expressions for the various $a$ and $b$.
The relic abundance is then obtained as:
   \begin{equation}
   \Omega_\chi h^2 \approx \frac{1.07 \times 10^9~\text{GeV}^{-1} }{M_\text{Pl}}\frac{x_F}{\sqrt{g_*}} \frac{1}{I_a+3I_b/x_F}  ~,
   \label{eq:relabund}
   \end{equation}
where $M_{\rm Pl}$ is the Planck mass, $g_*$ is the number of relativistic degrees of freedom and the freezeout value of $x$ is solved from
   \begin{equation}
\nn   e^{x_F} = \frac{5}{4}\sqrt{\frac{45}{8}}\frac{\mdm M_{\text{Pl}}(I_a + 6I_b/x_F)}{\pi^3\sqrt{g_*}\sqrt{x_F}} ~.
   \end{equation}
$I_a$ and $I_b$ capture the thermal history
        of DM before freezeout:
   \bea
  \nn I_a &=& x_F (1+\delta)^3 \int^\infty_{x_F} \frac{dx}{x^2}~ e^{-2\delta x}~a~,\\
    I_b &=& 2 x_F^2 (1+\delta)^3 \int^\infty_{x_F} \frac{dx}{x^3}~e^{-2\delta x}~b~.
   \label{eq:thermalhistory}
   \eea

We end this section with two remarks. 
First, we comment on the validity of the Taylor expansion $a + b~v^2$, 
where $v$ is the relative velocity of a top quark pair. 
The cross-section times velocity,  $\sigma_{t \bar{t} \ra \dm \dm} v$, carries a phase space factor $|\vec{p}_\dm|/\mtop$, and $|\vec{p}_\dm|^2 = (\mtop^2-\mdm^2) + \mtop^2v^2/4$.
The expansion in $v$ hence assumes $(\mtop^2-\mdm^2) >   \mtop^2v^2/4$, or $\delta (2+\delta)/(1+\delta) > v^2/4$.
At the time of freezeout, $v \simeq 0.3$, which means the expansion breaks down for $\delta \lsim 0.025$.
This translates to $\mdm > 170$~GeV, and for this reason, we present constraints only for $\mdm \leq 170$~GeV.

Second, we remark on the difference between the treatment leading to Eq.~(\ref{eq:thermalXSmaster}) for forbidden DM and the usual procedure used in DM annihilation.
DM is usually taken to annihilate to relativistic species, in which case Eq.~(\ref{eq:thermalXSmaster}) is not helpful. 
Specifically, the thermal distribution of $\psi_i$ cannot be approximated by Maxwell-Boltzmann statistics, and Fermi-Dirac or Bose-Einstein statistics must be used instead, complicating the calculation of $\sigmavee_{\psi_1 \psi_2 \ra \chi_1 \chi_2}$.
Therefore, the Boltzmann equation is simplified using detailed balance and unitarity in a way that eliminates all information about the annihilation products $\psi_i$:
\beq
\nn \dot{n}_{\chi_{1,2}} + 3 H n_{\chi_{1,2}} = - \sigmaveeave (n_{\chi_{1}}n_{\chi_{2}}-n^{\rm eq}_{\chi_{1}}n^{\rm eq}_{\chi_{2}})~.
\eeq

From here the usual approximations of  \cite{Gondolo:1990dk} proceed and the relic density is obtained.

\section{Formulae}
\label{app:formulae}

We provide in this appendix formulae for the calculation of the thermal 
and direct detection cross-sections. 
The formulae for $a$ and $b$ in the following (as given by $\sigmaveeave = a + b v^2 + \mathcal{O}(v^4)$) are to be used in Eq.~(\ref{eq:relabund}) in order to obtain the forbidden relic abundance.
The spin-independent direct detection cross-section for dark matter-nucleon scattering is given by 
\beq
\sigma_{\dm N} = \frac{4}{\pi} \mu_N^2 |f_N|^2~,
\eeq
where $\mu_N$ is the $\dm$-nucleon reduced mass and $f_N$ the effective coupling
obtained from nucleon matrix elements of gluon operators. 
The expressions for $f_N$ for the various dark matter candidates considered are
provided below.
In the following, the number of QCD colors $N_c$ = 3, the mass of the top quark is taken
$\mtop$ = 174~GeV and $r \equiv (\mdm/\mtop)^2.$

\subsection{$s$-channel mediator: Dirac dark matter}

The Lagrangian (in 4-component notation) in the broken electroweak phase is given by
\bea
\nn \mathcal{L} \supset &-& \mdm \bar{\dm} \dm - \half \mmed^2 \smed^2  \\
&-& (i \ldm \smed \bar{\dm} \gamma_5 \dm + i \lsm \smed \bar{t} \gamma_5 t + {\rm h.c.})~.
\eea

Observe that the interaction of the mediator with top quarks might arise in the unbroken phase from a term such as

\begin{equation*}
\mathcal{L} \supset \frac{c}{\Lambda} H \phi~\bar{Q}_3  t^c~ + {\rm h.c.}~,
\end{equation*}

where $\phi$ is a complex scalar containing the pseudoscalar mediator $\smed$.

\subsubsection{\bf\em Relic density}

The Taylor co-efficients of  $\sigmavee_{t \bar{t} \ra \dm \dm}$ to be used in Eq.~(\ref{eq:thermalhistory}) are given by

\bea
\nn a = \frac{c_T}{2\pi N_c}\frac{\lambda^4\mtop^2\sqrt{1-r}}{(\mmed^2-4\mtop^2)^2+\decay^2\mmed^2}~, \\
\nn b = \frac{c_T}{16\pi N_c}\frac{\lambda^4\mtop^2}{(\mmed^2-4\mtop^2)^3+\decay^3\mmed^3} \times \\
\frac{\mmed^2(2-r)+4\mtop^2(2-3r)}{\sqrt{1-r}}~.
\label{eq:s-chsigmav}
\eea
The factor $c_T = 1/2$ for Dirac $\chi$, coming from the thermal averaging.
If we had chosen $\chi$ Majorana instead, $c_T = 1$.

\subsection{$t$-channel mediator: fermionic dark matter}

Assuming $\dm$ to be a Majorana fermion, the masses and relevant interactions are given by 
\bea
\nn \mathcal{L} \supset &-& \half\mdm(\dm^2+{\dm^\dagger}^2) - \mmed^2 |\med|^2  \\
&-& (\lambda \med^* \dm t_R  + {\rm h.c.}) 
\eea

\subsubsection{\bf\em Relic density}

\bea
\nn a = \frac{\lambda^4\mtop^2 r \sqrt{1-r}}{32\pi N_c (\mmed^2+\mtop^2-\mdm^2)^2}~, \\
\nn b = \frac{ \lambda^4 \mtop^2 }{768\pi N_c (\mmed^2+\mtop^2-\mdm^2)^4\sqrt{1-r}} \\
\nn \times [ -2\mmed^2\mtop^2 r (22-35r+13r^2) \\
\nn + \mmed^4(16-26r+13r^2) \\
 +\mtop^4(r-1)^2(16-10r+13r^2)]~.
\label{eq:Majsigmav}
\eea
We find Eq.~(\ref{eq:Majsigmav}) in agreement with \cite{Chang:2013oia}.

\subsubsection{\bf\em Direct detection}

Following \cite{Hisano:2010ct}, we obtain the effective coupling as 
\beq
f_N = - m_N \frac{8\pi}{9} f_{TG} f_G
\eeq
where $m_N$ is the mass of the nucleon, $f_{TG}$ is the mass fraction of
the gluon with values (taken from \cite{Hisano:2010ct})
\beq
f_{TG}^{\rm proton} = 0.925~, ~~f_{TG}^{\rm neutron} = 0.922~
\label{eq:fTGs}
\eeq
and $f_G$ is the effective $\dm$-gluon
coupling obtained by evaluating the gluon loop. 
It is separated into short- and long-distance effects given as
\beq
f_G = \frac{\lambda^2}{32\pi} \mdm (f_{\rm SD} + c_t f_{\rm LD}), 
\eeq
where $c_t = 1 + 11\alpha_s(\mtop)/4\pi$ is the QCD correction, taken as 1.
$f_{\rm SD}$ and $f_{\rm LD}$ are evaluated as 
\bea
\nn f_{\rm SD} &=&  - \frac{(\Delta - 6\mmed^2\mtop^2)(\mmed^2+\mtop^2-\mdm^2)}{6\Delta^2\mmed^2} \\
\nn && - \frac{2\mmed^2\mtop^4}{\Delta^2}L~, \\
f_{\rm LD} &=& -\frac{\Delta+12\mmed^2\mtop^2}{6\Delta^2}\\
\nn &&+ \frac{\mmed^2\mtop^2(\mmed^2+\mtop^2-\mdm^2)}{\Delta^2}L~,
\eea
with
\beq
\Delta = \mdm^4 - 2\mdm^2(\mmed^2+\mtop^2)+(\mmed^2-\mtop^2)^2
\eeq
and
\beq
 L =
    \begin{cases}
      \frac{1}{\sqrt{\Delta}} \log\left(\frac{\mmed^2+\mtop^2-\mdm^2+\sqrt{\Delta}}{\mmed^2+\mtop^2-\mdm^2-\sqrt{\Delta}}\right)&,~ \Delta>0~, \\
       \frac{2}{\sqrt{\Delta}} \tan^{-1}\left(\frac{\sqrt{|\Delta|}}{\mmed^2+\mtop^2-\mdm^2}\right)&,~\Delta<0~.
    \end{cases}
\eeq

\subsection{$t$-channel mediator: scalar dark matter}

Taking $\dm$ to be a real scalar, the relevant Lagrangian is

\bea
\nn \mathcal{L} \supset &-& \half \mdm^2 \dm^2 - \left(\half \mmed \med^2  
+ \lambda \dm \med t_R  + {\rm h.c.}\right) 
\eea
\\

\subsubsection{\bf\em Relic density}
\bea
\nn a = \frac{\lambda^4 \mtop^2 (1-r)^{3/2}}{16\pi N_c (\mmed^2+\mtop^2-\mdm^2)^2}~, \\
\nn b = \frac{\lambda^4 \mtop^2 \sqrt{1-r}}{384\pi N_c(\mmed^2+\mtop^2-\mdm^2)^4} \\
\nn \times [ -2\mmed^2\mtop^2  (r-1)(r-8) \\
\nn +\mtop^4(r-1)^2(r-8)  \\
+ \mmed^4(r+8)]~.
\label{eq:scalsigmav}
\eea

\subsubsection{\bf\em Direct detection} 
To our knowledge, an explicit loop calculation involving the top quark mass  
has not been performed in the literature, resulting in an overestimation
of the scattering cross-section. 
The effective coupling is given by \cite{Hisano:2015bma}
\beq
f_N =  - \left(\frac{m_N}{2\mdm}\right) \frac{8\pi}{9} f_{TG} f_G~,
\eeq
\newline

where the Wilson coefficient is
\beq
f_G =  - \frac{\lambda^2}{24(\mmed^2-\mdm^2)}~
\eeq
and the $f_{TG}$'s are given in Eq.~(\ref{eq:fTGs}).



\end{document}